\documentclass[aps,showpacs,twocolumn,10pt,prl]{revtex4-1}

\usepackage{amsmath}
\usepackage{amsthm}
\usepackage{amsfonts}
\usepackage{amssymb}
\usepackage{graphicx}

\begin{document}
\title{Electronic dynamics and frequency-dependent effects in circularly polarized strong-field physics}

\author{F. Mauger}
\author{A.~D. Bandrauk}
\affiliation{Laboratoire de Chimie Th\'eorique, Facult\'e des Sciences, Universit\'e de Sherbrooke, Sherbrooke, Qu\'ebec, Canada J1K 2R1}
\date{\today}

\begin{abstract}
We analyze, quantum mechanically, the dynamics of ionization with a strong, circularly polarized, laser field. We show that the main source for non-adiabatic effects is connected to an effective barrier lowering due to the laser frequency. Such non-adiabatic effects manifest themselves through ionization rates and yields that depart up to more than one order of magnitude from a static-field configuration. Beyond circular polarization, these results show the limits of standard instantaneous  -- static-field like -- interpretation of laser-matter interaction and the great need for including time dependent electronic dynamics.
\end{abstract}
\maketitle


Strong-field physics, which analyzes the interaction between very strong ultra-short laser pulses and atoms or molecules, attracts an increasing interest as new tools to probe matter at the spatio-temporal resolution of the electronic dynamics in these systems in the nonlinear non-perturbative regime~\cite{Beck08,Mori11,Beck12}. The intrinsic coherence of the recollision mechanisms~\cite{Cork93,Scha93} now allows for tomographic orbital imaging~\cite{Itat04,Haes10,Cork11} and promises for retrieving spatial geometrical properties~\cite{Zuo96,Meck08,Pete11} of molecules: Upon electronic recollision, an instantaneous snapshot of quantum structures is imprinted in the sub-products of the laser-matter interaction. Yet, since the laser has the ability to distort orbitals, the question of what is imaged exactly remains. With linear polarization, this difficulty is often eluded by arguing that recollision, i.e., the time at which the imaging occurs, mostly corresponds to a zero of the field~\cite{Itat04} and unperturbed orbitals are retrieved. In this Letter, using the specific example of circular polarization (CP) excitation ionization, we show that one cannot disregard dynamical effects due to the overall interaction with the laser and that these effects manifest themselves through frequency-dependent non-adiabatic effects.

A large body of literature has been dedicated to the computation of strong-field ionization rates, following the seminal works of Keldysh~\cite{Keld65}, Perelomov-Popov-Terentev~\cite{Pere66}, Ammosov-Delone-Krainov~\cite{Ammo86} and later including empirical corrections guided by numerical integration of the time-dependent Schr\"{o}dinger equation (TDSE)~\cite{Baue99,Tong05}. For typical experimental configurations, the difficulty for defining ionization rates comes from the constantly changing amplitude and/or direction of the laser field. Following the aforementioned instantaneous picture of the laser effect, a usual shortcut consists of an adiabatic approximation and to consider, at each time, as if the system were subjected to a static field. Experiments with near circular polarization and Ar have revealed ionization statistics~\cite{Hera12} incompatible with such an adiabatic picture~\cite{Bart11} while other experiments with He tend to validate it~\cite{Boge13}. In this Letter, we reconcile these two results by showing that non-adiabatic effects are indeed at play but might be too weak to be observed with He.

\begin{figure} 
	\centering
		\includegraphics[width=\linewidth]{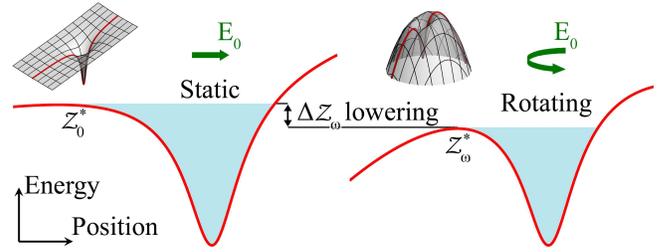}
	\caption{\label{fig:barrier_lowering}
	(color online) Barrier lowering induced by a CP laser field. Insets display the ZVS Eq.~(\ref{eq:zero_velocity_surface}) for static (left) and rotating (right) frames. Main panels show the corresponding section $\tilde{y}=0$ (thick red curve on the insets) passing by the ionization barrier (saddle, see text).}
\end{figure}

Among all laser polarizations, CP is unique in that the amplitude of the laser is constant and only its direction changes with time. As a consequence, an adiabatic dynamics would result in ionization rates and yields that do not depend on the laser frequency. Far from that, in this Letter, we make use of this symmetry to identify real laser-induced non-adiabatic effects, which are defined as any deviation from a purely static (DC) field with similar amplitude. More specifically, we show that Coriolis effects associated with the laser rotational dynamics induces an effective barrier lowering, as illustrated in Fig.~\ref{fig:barrier_lowering}, and we study its impact on ionization.

In the dipole approximation, and using atomic units (unless otherwise specified), the Hamiltonian operator corresponding to the interaction between an atomic or molecular system and an external laser field reads
\begin{equation} \label{eq:Hamiltonian_operator}
	\hat{\mathcal{H}} = -\frac{\Delta}{2} + \mathcal{V}\left({\bf x}\right)
		+ f\left(t\right) \frac{E_{0}}{\sqrt{1+\varepsilon^{2}}}\left( \cos\omega t\ \hat{x} + \varepsilon \sin\omega t\ \hat{y} \right),
\end{equation}
where $f$ is the laser envelope, $I_0\propto{E_{0}^{2}}$ its intensity with ellipticity $\varepsilon$ ($0\leq\varepsilon\leq1$ spans linear to circular polarization) and frequency $\omega$. In this Letter, we consider rotationally symmetric potentials $\mathcal{V}\left({\bf x}\right)=V\left(\left|{\bf x}\right|\right)$, where $\left|\cdot\right|$ is the Euclidean norm. We use soft-Coulomb potentials~\cite{Java88} to model atomic systems in reduced two dimensions $V\left(\left|{\bf x}\right|\right) = -Z_{\rm eff}/\sqrt{\left|{\bf x}\right|^{2} + a^{2}}$ where the effective charge $Z_{\rm eff}$ and softening parameter $a$ are chosen such as to reproduce the appropriate eigenstate in the free-field configuration~\cite{NIST_AtomicData,Note1}. A quick glance at Eq.~(\ref{eq:Hamiltonian_operator}) reveals two possible sources for dynamical effects in the associated electronic evolution, each one with a different time-scale. The laser envelope $f$ defines the absolute pulse duration and therefore corresponds to the ``slow'' time-scale; comparatively, the laser frequency defines a ``fast'' time-scale. In what follows, we investigate the respective role of each time-scale on the ionization dynamics and show that the main source for non-adiabatic effects is frequency related.

\begin{figure} 
	\centering
		\includegraphics[width=\linewidth]{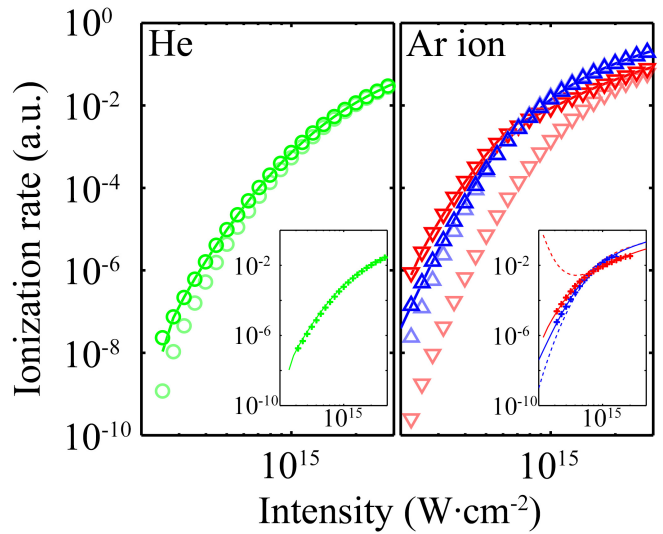}
	\caption{\label{fig:ionization_rate}
	(color online) 
	Main panels: Ionization rate for He (left) and Ar ion (right) models. In each panel, dark (resp. light) markers correspond to $800~{\rm nm}$ wavelength (resp. static field) and curves draw the ionization rate given by Eq.~(\ref{eq:ionization_rate}). Circles (green online) correspond to $s$ initial state, triangles (blue) correspond to $p_{x}$ (connected to positive angular momentum, $m=1$, see text) initial state while inverted triangles (red) correspond to $p_{y}$ (connected to negative, $m=-1$, angular momentum) initial state.
	Insets: Ionization rate fit curves. Dashed curves correspond to Eq.~(\ref{eq:ionization_rate}) and continuous curves to the best fit between low and high intensity regimes fits (see text). For comparison, markers show the ionization rates directly computed from numerical integration of the TDSE with Hamiltonian~(\ref{eq:Hamiltonian_operator}).}
\end{figure}

For the sake of simplicity, we treat separately the analysis of the two time-scales and begin with the ``fast'' one, associated with the laser frequency. For this purpose, we consider the idealistic situation of a constant envelope ($f=1$) such that all effects related to the ``slow'' time-scale are canceled. For a CP laser field ($\varepsilon=1$), moving the system into a frame rotating with the laser field maps the Hamiltonian~(\ref{eq:Hamiltonian_operator}) to
\begin{equation} \label{eq:Hamiltonian_operator_rotating_frame}
	\hat{\tilde{\mathcal{H}}} = -\frac{\Delta}{2} + \mathcal{V}\left(\tilde{\bf x}\right) 
		- \omega \hat{\tilde{\mathcal{L}}}_{z} + \frac{E_{0}}{\sqrt{2}}\hat{\tilde{x}},
\end{equation}
where tildes stand for rotating frame coordinates, and $\hat{\tilde{\mathcal{L}}}_{z}=-i\left(\hat{\tilde{x}}\partial_{\tilde{y}}-\hat{\tilde{y}}\partial_{\hat{\tilde{x}}}\right)$ is the angular momentum. The price to pay for moving the analysis into the rotating frame is the introduction of a Coriolis effect ($-\omega\hat{\tilde{\mathcal{L}}}_{z}$) but it has the invaluable advantage of removing all explicit time dependence and make the Hamiltonian operator autonomous. In such a configuration, from the theoretical point of view, ionization rates can be defined rigorously and are related to specific generalized eigenstates with complex energy $E_{r}=E-i\Gamma/2$ called resonances~\cite{Rein82}. The scalar $E$ is usually referred to as the energy of the resonance while $\Gamma$ ($\geq0$) is its width or ionization rate. Indeed, for a resonance $\left|\psi_{r}\right\rangle$, combining the generalized eigenstate relation $\hat{\tilde{\mathcal{H}}}\left|\psi_{r}\right\rangle=E_{r}\left|\psi_{r}\right\rangle$ with the Schr\"{o}dinger equation leads to an exponential decrease of the electronic density with rate $\Gamma$. In this Letter, all resonances are computed numerically through a partial diagonalization of a discretized representation of the Hamiltonian operator~(\ref{eq:Hamiltonian_operator_rotating_frame}) using complex coordinates~\cite{Rein82} and a thick restart Arnoldi strategy~\cite{Wu00}. Resonances are found by continuously varying the laser parameters ($E_{0}$ and $\omega$) from the relevant free-field eigenstates. Ionization rates are further confirmed by numerical integration of the TDSE, in the static frame, using the Hamiltonian operator~(\ref{eq:Hamiltonian_operator}) (see insets of Fig.~\ref{fig:ionization_rate})~\cite{Note2}.

Hamiltonian~(\ref{eq:Hamiltonian_operator_rotating_frame}) corresponds to the free-field problem to which are added the laser frequency and amplitude effects, respectively parametrized by $\omega$ and $E_{0}$. It explains how resonances can be followed from free-field eigenstates in the limit $\omega\rightarrow0$ and $E_{0}\rightarrow0$. Besides, for a fixed laser frequency, i.e., a given laser source, two intensity regimes are expected in the ionization dynamics: In the low intensity regime ($E_{0}\ll\omega$) Coriolis effects dominate while in the high intensity regime ($E_{0}\gg\omega$) static field-like effects take the lead, as we shall see in what follows. The transition between the two regimes depends on the target and initial state, through the principal (state symmetry, $n$) and magnetic (angular momentum, $m$) quantum numbers. We attribute those effects to strong non-adiabatic manifestations in the electronic dynamics, which cannot be neglected in standard experimental setups.

We compare, in Fig.~\ref{fig:ionization_rate}, the ionization rates for He ($s$ state, zero initial angular momentum $m=0$) and Ar ion ($p$ state, nonzero initial angular momentum) models for static and $800~{\rm nm}$ wavelength with CP. Non-adiabatic effects manifest themselves through ionization rates that depart up to more than one order of magnitude from static-field results, in the low intensity regime. We attribute these overwhelming effects to an effective barrier lowering induced by the rotational motion associated with CP (see Fig.~\ref{fig:barrier_lowering}). This effect is best understood considering the classical counterpart to Eq.~(\ref{eq:Hamiltonian_operator_rotating_frame}). In the rotating frame, the classical Hamiltonian reads~\cite{Maug10_3,Barr11}
$$
	\tilde{\mathcal{H}}\left(\tilde{\bf x}, \tilde{\bf p}\right) = \frac{\left|\tilde{\bf p}\right|^{2}}{2} 
		+ \mathcal{V} \left(\tilde{\bf x}\right) - \omega\tilde{\mathcal{L}}_{z} + \frac{E_{0}}{\sqrt{2}}\tilde{x},
$$
where $\tilde{\bf x}$ and $\tilde{\bf p}$ are canonically conjugated position and momentum, and the angular momentum $\tilde{\mathcal{L}}_{z}=\tilde{x}\tilde{p}_{y}-\tilde{y}\tilde{p}_{x}$. From there, the limits of the classically accessible domain to the electron dynamics are charted by the so-called zero-velocity surface (ZVS)~\cite{Hill78,Maug10_3,Barr11} of equation
\begin{equation} \label{eq:zero_velocity_surface}
	\mathcal{Z}_{\omega}\left(\tilde{\bf x}\right) = 
		- \frac{\omega^{2}}{2}\left|\tilde{\bf x}\right|^{2} 
		+ \mathcal{V}\left(\tilde{\bf x}\right) + \frac{E_{0}}{\sqrt{2}}\tilde{x},
\end{equation}
which is deduced from the classical Hamiltonian through a non-canonical change of variables to position-velocity coordinates [$\mathcal{Z}_{\omega}\left(\tilde{\bf x}\right)=\mathcal{H}\left(\tilde{\bf x},\dot{\tilde{\bf x}}={\bf 0}\right)$]. We see that the laser frequency induces a barrier lowering ($-\omega^{2}\left|\tilde{\bf x}\right|^{2}/2$ term), increasing with the frequency, irrespective of the polarization direction, left or right: Here, the rotational symmetry of the potential is incompatible with circular dichroism -- defined as an asymmetry between left and right polarizations -- and that, here, can only arise from dynamical effects, e.g., through the initial state configuration. More precisely, the ionization barrier is defined by a saddle point, in phase space~\cite{Maug10_3,Barr11}, whose position corresponds to a saddle on the ZVS~\cite{Hill78}. We define $\mathcal{Z}^{*}_{\omega}$ as the energy of the ionization barrier and $\Delta\mathcal{Z}_{\omega}=\mathcal{Z}^{*}_{0}-\mathcal{Z}^{*}_{\omega}$ the barrier lowering induced by the laser frequency, compared to the static field configuration $\omega=0$ (see Fig.~\ref{fig:barrier_lowering}). Numerical analysis of reduced two-dimensional models reveals that non-adiabatic effects can be factorized into a correction to static field ionization rates
\begin{equation} \label{eq:ionization_rate}
	\Gamma_{\omega}\left(E_{0}\right) \approx \Gamma_{0}\left(E_{0}\right) \exp\left( \beta \Delta\mathcal{Z}_{\omega}^{\alpha}\right),
\end{equation}
for some constants $\alpha$ and $\beta$ that depend on the intensity regime (low intensity, Coriolis effect dominating, or  high intentity, static-field like) and on the initial state (quantum numbers), i.e., the atomic properties, for each species. The simplicity of formula~(\ref{eq:ionization_rate}) along with its robustness as the wavelength and target species are varied~\cite{Note3} show the central role played by the barrier lowering effect associated with the CP induced rotational electronic dynamics. The exponential dependence, further amplified by the fact that ionization yields themselves depend exponentially on the rates, is at the heart of frequency induced non-adiabatic effects in strong field physics.

In order to substantiate the factorized expression of formula~(\ref{eq:ionization_rate}), we begin the analysis with the simplest configuration of a symmetric ($s$) initial state and take a He model atom. This state is non-degenerate and has zero angular momentum ($m=0$). Therefore, looking at Hamiltonian~(\ref{eq:Hamiltonian_operator_rotating_frame}) no circular dichroism in the ionization rate is expected. Numerical simulations confirm that the ionization rate only depends on the magnitude of the laser frequency $\omega$ and field amplitude $E_{0}$. In this configuration, analysis of the ionization rates shows that the barrier lowering effect enhances ionization for all intensities and the difference gets more pronounced for low intensities (see markers in the left panel of Fig.~\ref{fig:ionization_rate}). At $800~{\rm nm}$, the fitting parameters in Eq.~(\ref{eq:ionization_rate}), obtained through a linear regression of ionization rates deduced from the resonances, are $\alpha=3.17$ and $\ln\left(\beta\right)=11.7$ and yield almost a perfect match with a direct computation of the ionization rate (compare markers and the curve in the Fig.). Further analysis reveals that, for a fixed laser intensity, non-adiabatic effects get stronger when the laser frequency (resp. wavelength) is increased (resp. decreased), as can be expected from Eq.~(\ref{eq:Hamiltonian_operator_rotating_frame}).

We now consider the situation of a degenerate $p$ state with non-zero angular momentum and consider Ar ion model. In this situation, Hamiltonian~(\ref{eq:Hamiltonian_operator_rotating_frame}) shows that the polarization direction, i.e., the sign of $\omega$, matters in the ionization dynamics and circular dichroism is expected in ionization rates. Equation~(\ref{eq:ionization_rate}) suggests comparison of the ionization process with the static field situation. Static fields lift the state degeneracy and an aligned ($p_{x}$) orbital is preferred compared to the orthogonal ($p_{y}$) state configuration for ionization (see light markers in the right panel of Fig.~\ref{fig:ionization_rate}). Turning on the laser frequency shows that negative, $m=-1$, (resp. positive, $m=1$,) angular momentum is connected to the $p_{y}$ (resp. $p_{x}$) eigenstate: Examining the corresponding generalized eigenstates shows that they change their shape to converge towards resonances with the identified orientation ($x$- or $y$- alignment). As introduced previously, the low intensity regime is dominated by Coriolis effects such that the barrier lowering induces ionization rate enhancement, irrespective of the angular momentum. However, the effect is far more pronounced when the Coriolis effect and natural electron rotation coincide, i.e., for negative angular momentum ($m=-1$, with $\omega>0$), than it is with opposite effects, leading to higher ionization rates (see dark markers)~\cite{Hera12,Bart11}. At $800~{\rm nm}$, the fitting parameters in Eq.~(\ref{eq:ionization_rate}) are $\alpha=1.66$ and $\ln\left(\beta\right)=6.56$ for negative angular momentum ($m=-1$) and $\alpha=5.23$ and $\ln\left(\beta\right)=14.6$ for positive ($m=1$) one. On the other hand, the high intensity regime is dominated by static-field like effects, such that ionization rate curves intersect and reverse their order. For $p_{y}$ states (connected to negative angular momentum, $m=-1$) the laser induced and natural electron rotation still coincide, leading to a favorable ionization situation and thus a positive $\beta$ parameter. At $800~{\rm nm}$, we find $\alpha=3.69$ and $\ln\left(\beta\right)=12.98$ as a best fit to Eq.~(\ref{eq:ionization_rate}). On the contrary, for a $p_{x}$ state (connected to positive angular momentum, $m=1$) the two rotational effects counteract, leading to a slightly smaller ionization rate and thus negative $\beta$ parameter. At $800~{\rm nm}$, we find $\alpha=3.41$ and $\ln\left(-\beta\right)=9.39$ as best fit to Eq.~(\ref{eq:ionization_rate}). Comparing markers and their corresponding curves on the right panel of Fig.~\ref{fig:ionization_rate}, we notice the good agreement with the result of Eq.~(\ref{eq:ionization_rate}) both in low and high intensity regimes. Similarly to the initial $s$ state configuration, further analysis shows that all the aforementioned effects get more pronounced as the laser frequency is increased, in agreement with Hamiltonian~(\ref{eq:Hamiltonian_operator_rotating_frame}). Finally, as can be expected from the expression of Hamiltonian~(\ref{eq:Hamiltonian_operator_rotating_frame}), the transition between low intensity regime, dominated by Coriolis effects, and high intensity regime, where a static-field like configuration takes the lead, shifts towards higher intensities when the laser frequency is increased. To conclude, the present ``fast'' time-scale analysis clearly shows that strong, frequency-dependent, non-adiabatic effects are commonly at play in strong-field physics.

\begin{figure} 
	\centering
		\includegraphics[width=\linewidth]{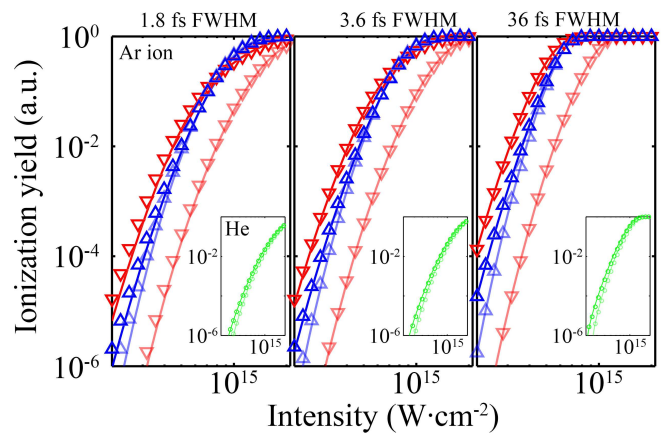}
	\caption{\label{fig:envelope_yield}
	(color online) Ionization yield for cosine square envelope for Ar ion (main panels, blue and red data online) and He (insets, green) models. Marker code (color and shade) is the same as in Fig.~\ref{fig:ionization_rate} and corresponds to numerical integration of the TDSE with Hamiltonian~(\ref{eq:Hamiltonian_operator}). For comparison, curves correspond to the ionization yields of Eq.~(\ref{eq:ionization_probability}).}
\end{figure}

Finally, we consider the ``slow'' time-scale and reintroduce the envelope $f\left(t\right)$. The change of coordinates into the rotating frame yields a non-autonomous system [the amplitude $E_{0}/\sqrt{2}$ is replaced with $f\left(t\right)E_{0}/\sqrt{2}$ in Hamiltonian~(\ref{eq:Hamiltonian_operator_rotating_frame})] and the rigorous ionization rate definition, based on resonances, breaks down. Yet, the slow variation of the envelope advocates for an adiabatic treatment where the instantaneous effective ionization rate is defined as the one with corresponding laser frequency and amplitude $\Gamma\left(t\right)=\Gamma_{\omega}\left(f\left(t\right)E_{0}\right)$. Then, neglecting the probability for recapture, the ionization probability is solution of $\dot{\mathbb{P}}\left(t\right)=\left(1-\mathbb{P}\left(t\right)\right)\Gamma\left(t\right)$ leading to
\begin{equation} \label{eq:ionization_probability}
	\mathbb{P}\left(t\right) = 1 - \exp\left(-\int_{-\infty}^{t}{\Gamma\left(s\right)\ ds}\right),
\end{equation}
assuming zero initial ionization. \emph{It is important to note that, here, the adiabatic approximation is taken on the envelope solely while frequency-dependent non-adiabatic effects are fully included.} In Fig.~\ref{fig:envelope_yield}, we compare ionization yields with a cosine square envelope and various pulse durations to direct integration of the TDSE with Hamiltonian~(\ref{eq:Hamiltonian_operator}). Overall we notice the very good agreement with Eq.~(\ref{eq:ionization_probability}); only for the shortest pulses do the results slightly depart. For He, frequency-dependent non-adiabatic manifestations, in the near infrared regime, are limited (see insets) such that they might not be accessible to current experimental setups~\cite{Boge13}. On the other hand, for Ar ion, because of the non-zero angular momentum, non-adiabatic effects associated with the laser frequency play a dramatic role, eventually leading to orders of magnitude differences in the ionization yields (compare dark and light data) and eventually leading to experimentally observable manifestations~\cite{Hera12}.

In summary, we have shown that non-adiabatic effects, mostly due to the laser frequency, are responsible for dramatic changes in ionization rates and yields with CP. It clearly shows that for typical experimental configurations -- near infrared laser, rare gas target -- an instantaneous representation of the quantum mechanical system, as if the laser field is static, is limited. Dynamical effects resulting from the overall interaction with the laser cannot be disregarded. Beyond CP, our results call for further and more careful analysis of laser-matter interaction, both from the theoretical and experimental points of view, and that for all polarizations, including linear as is commonly considered in strong-field physics.

%
F.M.\ acknowledges enlightening discussions with E.~Soccorsi and O.~Atabek.
The authors thank RQCHP and Compute Canada for access to massively parallel computer clusters and the CIPI for financial support in its ultrafast science program.
The authors acknowledge financial support from the Centre de Recherches Math\'ematiques. F.M.\ acknowledges financial support from the Merit Scholarship Program for Foreign Student from the MESRS of Quebec. A.D.B.\ acknowledges financial support from the Canada Research Chair.




\end{document}